\documentclass[a4paper,11pt]{article}
\usepackage{jheppub}
\usepackage{lineno}
\usepackage{subcaption}

\title{Analytical approximations as close as desired to special functions}
\author{Aviv Orly}
\affiliation{School of Physics and Astronomy\\
Tel Aviv University\\ 6997801 Ramat Aviv, Israel}

\emailAdd{avivbinyaminorly@gmail.com}

\abstract{We present a method for constructing global analytical expressions that approximate a function over its entire range. These approximations not only mirror the original function as accurately as desired, but are purposely created to possess features that the original function lacks. This is particularly useful for functions that lack closed form and are defined by integrals or infinite series. Replacing these definitions with simple analytical expressions enables in-depth qualitative analysis and replaces the current methods of evaluation. We demonstrate this procedure by providing replacements for a variety of pivotal functions in physics and cosmology including the pressure and density of quantum gas, the one-loop correction in thermal field theory, common polylog functions, and the error function.}

\begin{document}
\maketitle
\flushbottom
\section{Introduction}
\label{sec:intro}
Many kinds of functions in research do not behave as we want them to. Among the many desired properties are integrability, differentiability, invertibility, evaluation complexity, simplification by algebraic manipulation with other functions, and the existence of closed-form expression. We develop a methodology that enables crafting an approximate analytical function that possesses desired qualities, which is as close as desired to the original. These analytical terms alias the objective function for any practical consideration, thus suppressing the unwanted nature of the original function to remain theoretical only.

This paper demonstrates this approach on non-elementary functions relevant to cosmology. Non-elementary functions are functions that a finite number of elementary functions cannot express. Instead, they are defined as integrals, infinite sums or solutions to differential equations. The absence of a closed form can prevent the possibility of qualitative analysis and, in many cases, causes evaluation challenges.

Although accurate evaluation of these functions is possible, it often presents a challenging task that is not directly related to primary research. For example, evaluating the pressure and density of Fermi gas is done by various integrations over the Fermi-Dirac statistics. These integrations must be done numerically in the general case\cite{cosmology}. This can be an unwanted complication and an extremely inefficient or numerically unstable process. Asymptotic expansions are available only for certain limits. Therefore, an increasing number of terms is needed in the intermediate range to control the error. Crafting a single expression for the entire range not only makes the numerical integration unnecessary but enables some qualitative analysis of the entire range at once, which was not possible before. We focus on single variable functions, and touch on multivarible functions in section~\ref{sec:nonzero_chemical_potential}.

\section{Known approximation techniques}
The most widely known method for approximating a function near a given point is the Taylor series expansion\cite{arfken}. Naively, for approximating a function, $f(x)$ in a given range, one may consider a two-point Taylor expansion\cite{twopointtaylor} which enables interpolation between the points. A more robust structure is the Padé approximant\cite{padeSimpleBook}, which is defined by a rational function of a given order, i.e., a ratio of two polynomials. It is extensively used in computational mathematics and theoretical physics\cite{pozzi,prevost,bultheel,george1971pade,jordan}. Padé approximants often provide better approximations than Taylor series expansions and may still converge where the Taylor series does not. Thus, when the behavior of a function is known between two points, expansions are often interpolated using two-point Padé approximants\cite{baker1996pade,bender1999}. 

Taylor and Padé approximations are excellent at approximating near a point or at some subrange however they might diverge from the function elsewhere as they are not designed to replace the objective function in the entire range.

A known procedure for constructing a single analytic structure that approximates the objective function on the entire range is known as the method of matching asymptotic expansions\cite{matchingBook1,AsymptoticHistory}. In this method one expands the objective function in its desired range limits, then, by requiring that the different expansions match in the middle, a global approximation solution is formed. The limitation of this method is that the accuracy is not controllable.

We develop a methodology that combines the advantages from both approaches, enabling the approximations to both be valid over the entire range while being able to control the accuracy as well.

Related works include the work on best approximating polynomials\cite{bestPolynom}, the anharmonic oscillator analysis\cite{anharmonic_alex}, the MPQA method\cite{MARTINK0} and the uniform Padé generalization work\cite{winitzki2003uniform}.

\section{Methodology}
The objective is to craft an approximation $\tilde f(x)$ that is as close as desired to a given function, $f(x)$, in a chosen range that could possibly be the entire real axis. We do this by first obtaining the asymptotic expansions of $f(x)$ at both edges of the range, and then combining them into a single expression that maintains this asymptotic behavior at both edges. The combined expression is naturally only accurate near the edges, while the middle range accuracy still needs to be tamed. The middle range accuracy is controlled by adding additional terms that include additional degrees of freedom (DOF) in the form of coefficients that are determined by numerical minimization of the resultant error.

We denote the range over which $f(x)$ is approximated as $[x_i,x_f]$. 
First, we obtain the asymptotic behavior of $f(x)$ near the edges, i.e., finding the functions $f_i(x)$ and $f_f(x)$ that satisfy $\lim_{x\rightarrow x_{i}}\frac{f\left(x\right)}{f_{i}\left(x\right)}=1$ and $\lim_{x\rightarrow x_{f}}\frac{f\left(x\right)}{f_{f}\left(x\right)}=1$. These two asymptotics are then combined to form the initial structure $f_s(x)$ that satisfies 
\begin{equation}
\label{structureCondition}
    \lim_{x\rightarrow x_{i}}\frac{f\left(x\right)}{f_{s}\left(x\right)}=\lim_{x\rightarrow x_{f}}\frac{f\left(x\right)}{f_{s}\left(x\right)}=1\;.
\end{equation}
A simple recipe for constructing $f_s$ is by addition:
\begin{equation}
\label{f_s}
    f_s(x)=g_i(x)f_i(x)+g_f(x)f_f(x)\;,
\end{equation}
where $g_{i}(x)$, $g_{f}(x)$ are auxiliary functions that control the unwanted interference between $f_i(x)$ and $f_f(x)$ when it exists. Unwanted interference happens when $f_i(x)$ is dominant at $x_f$, i.e. $\lim_{x\rightarrow x_f}\frac{f_i(x)}{f_f(x)}\ne0$, or vice versa, thus preventing the condition of eq. \eqref{structureCondition}. Thus, we require $\lim_{x\rightarrow x_i}g_i(x)=\lim_{x\rightarrow x_f}g_f(x)=1$, which maintain the correct convergence, then determine the rest of them by the interference. When no interference exists, we set $g_{i}(x)=g_f(x)=1$. When interference is present we choose $g$ to decay from one to zero towards the other edge. As an example, $f_s(x)$ of \eqref{eq:basicmu0} has $g_i(x)=e^{-x}$ and $g_f(x)=1$, to fix the unwanted dominance of $f_i(x)$ over $f_s(x)$ at $x\rightarrow x_f$.

The next step is creating the final approximation $\tilde f(x)$ on the basis of $f_s(x)$. This is done by algebraic manipulations of $f_s(x)$ that incorporate additional DOF while maintaining the correct limiting behavior. The objective is to modify the structure to possess the desired properties while striking a balance between simplicity and accuracy. We see that power series  e.g.,  \eqref{eq:mu0lessonepercent}, and  Padé approximations e.g. eq. \eqref{eq:mu0decent}, are excellent candidates as the additional DOF.

Once that full approximation structure has been set, we proceed into numerical minimization of the parameters to control the accuracy. The relative error of an approximation is defined as:
\begin{equation}
\label{eq:err}
\mathcal{\varepsilon_{R}}(x) = \left| \frac{f(x) - \tilde{f}(x)}{f(x)} \right|\;.
\end{equation}
The global accuracy of an approximation ,$\varepsilon_{R}$, is given by its p-norm ($1\le p$):
\begin{equation}
\label{eq:pnorm}
    \| \varepsilon_{R} \|_p=\bigg(\int_{x_{i}}^{x_{f}}dx\left|\frac{\tilde{f}(x)-f(x)}{f(x)}\right|^{p}\bigg)^{\frac{1}{p}}\;.
\end{equation}
Minimizing $\| \varepsilon_{R} \|_p$ enables distributing accuracy efforts uniformly across the entire range. The larger $p$, the more accurate the balance is towards minimizing the maximal error.

Minimizing a multivariable function is a central challenge in machine learning. As there is no closed way to guarantee the discovery of the global minimum, many algorithms have been developed that search the parameter space after local minima from which the best one is chosen. To make the search as broad as possible we use both gradient and non gradient based methods, as either can be suitable for a given case. First, we use the widely used local minimization gradient descent method\cite{gradientdiscent}. To conduct a global search, the gradient decent is initiated from a great number of starting guesses chosen by the Monte Carlo method\cite{2020MonteCarlo}. For non gradient based minimization, we used the differential evolution algorithm\cite{differentialEvolution} which is a specialized global optimization technique that evolves a population of solutions by combining them in specific ways.

For efficient convergence, it is crucial to adopt a sophisticated scoring system, otherwise the minimization may converge impractically slowly or converge to local minima, missing the global one altogether, resulting in loss of accuracy. The current analysis utilized a system where \(p\) in eq. \eqref{eq:pnorm} began from $p=2$, which represents the standard root mean square, and was gradually taken to \(\infty\). Beginning from a large value of \(p\) significantly weakens the convergence as the larger \(p\) is, the weaker the gradient correlates to the global shape of the function but only to the point of the highest error. 

\section{Walk-through example: Fermi gas pressure}
As a walk-through example we choose to craft an approximation for the widely used integral
\begin{equation}
\label{eq:pressureIntegral}
f(\lambda,\nu)=\int_{1}^{\infty}dx\frac{\left(x^{2}-1\right)^{\frac{3}{2}}}{e^{\lambda x-\nu}+1}\;.
\end{equation}
It serves as a good example for several reasons. This integral cannot be evaluated analytically\cite{simplePRDexpansion,cosmology}. It notably represents the pressure of quantum Fermi gas\cite{cosmology},  and the one-loop thermal correction in quantum field theory\cite{thermalfunctionnumerical}, which are two important and repeatedly used quantities in physics. Useful expansions in limits of high and low temperature exist, however they break down in the transition, which has great physical importance.

The pressure of Fermi gas is given by $p=\frac{gT^{4}}{6\pi^{2}}\lambda^{4}\times f$, where $g$ is the gas degeneracy factor, $T$ is the temperature, $m$ is the mass, $\mu$ is the chemical potential and $\lambda\equiv\frac {m}{T}$, $\nu\equiv\frac {\mu}{T}$. 

The one-loop thermal correction in quantum field theory, is calculable in terms of the thermal function $J_F$\cite{thermalfunctionnumerical,J_in_PT1}:
\begin{equation}
J_F\left(\lambda\right) \equiv  \int_{0}^{\infty} dxx^{2}\ln\left(1 + e^{-\sqrt{x^{2}+\lambda^{2}}}\right) = \frac{\lambda^{4}}{3}f\left(\lambda,\mu=0\right)\;.
\end{equation}
This thermal function is especially important in phase transitions, therefore is relevant for cosmology for the research of phase transitions in the early universe and beyond standard model physics that might have been in play there\cite{beyondSMYann}.

We start from the case of $\mu=0$ by crafting an approximation model for the purpose of providing replacements for negligible chemical potential pressure calculation and the thermal function. Then we tackle the case of non negligible chemical potential, by replacing the integrand with an analytically integrable approximation. Generalization to bosons is found in section~\ref{BandFpressureIntegral}.
\subsection{Zero chemical potential}
When the chemical potential is either zero or totally negligible, we are left with:
\begin{equation}
\label{eq:pressureEquationZeromu}
   f(\lambda)=\int_{1}^{\infty}dx\frac{\left(x^{2}-1\right)^{\frac{3}{2}}}{e^{\lambda x}+1}\;.
\end{equation}

\begin{figure}[t!]
  \centering
  \begin{subfigure}{0.5\textwidth}
    \centering
    \includegraphics[width=\textwidth]{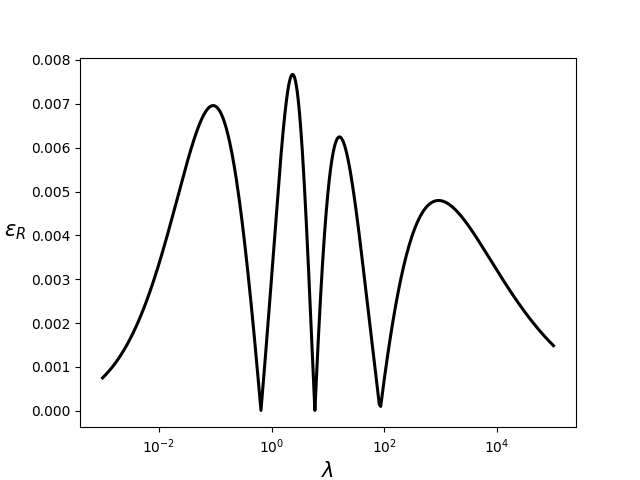}
    \caption{\(\varepsilon_R\) vs. \(\lambda\) for eq. \eqref{eq:mu0lessonepercent}}
    \label{fig:K0_1}
  \end{subfigure}%
  \begin{subfigure}{0.5\textwidth}
    \centering
    \includegraphics[width=\textwidth]{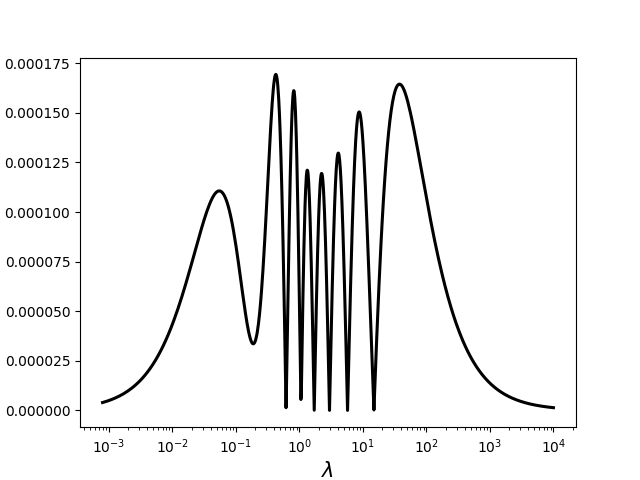}
        \caption{\(\varepsilon_R\) vs. \(\lambda\) for eq. \eqref{eq:mu0decent}}
    \label{fig:K0_2}
  \end{subfigure}
  \caption{Relative errors($\varepsilon_R$) as function of $\lambda$ for Fermion pressure approximations.}
  \label{fig:pressure_less_then_1}
\end{figure}

First, we analyze the asymptotic behavior of $f$, which is given by\cite{TheEarlyUiverse}:
\begin{equation}
\label{eq:lamdazero}
  f_i(\lambda)=f(\lambda\simeq0)=\frac{7\pi^4}{120\lambda^4}\;,
\end{equation}
\begin{equation}
    f_f(\lambda)=f(\lambda\simeq \infty)=\frac{3 \sqrt{\frac{\pi }{2}} e^{-\lambda }}{\lambda ^{5/2}}\;.
\end{equation}
Around zero, $f_f$ is negligible compared to $f_i$, i.e. $\lim_{\lambda \rightarrow 0}\frac{f_f(x)}{f_i(x)}=0$, as needed.
For large $\lambda$, $f_i$ is dominant, i.e. $\lim_{\lambda \rightarrow 0}\frac{f_i(x)}{f_f(x)}\ne0$, which creates unwanted overlap.
Therefore, to diminish the unwanted overlap we set $g_i(x)=e^{-\lambda}$ and $g_f(\lambda)=1$. Plugging that into eq. \eqref{f_s}, we get
\begin{equation}
\label{eq:basicmu0}
f_s=e^{-\lambda } \left(\frac{\sqrt{\frac{27 \pi }{2}}}{\lambda
   ^{5/2}}+\frac{7 \pi ^4}{120 \lambda ^4}\right)\;.
\end{equation}
This initial structure converges at both limits, with a highest relative error of around $0.37$ that occurs in the transition. We proceed by adding additional DOF in order to control the error. Adding $\lambda$ powers between $\frac52$ and $4$ keeps both asymptotic behavior intact while enabling us to reshape the transition range. By fitting both the coefficients and the powers of two added terms, as can be seen in figure~\ref{fig:pressure_less_then_1}, we improve the maximal relative error to less than of one percent, by the following:
\begin{equation}
\label{eq:mu0lessonepercent}
\tilde{f}=e^{-\lambda}\left(\frac{111.3}{\lambda^{3.05}}-\frac{108}{\lambda^{3.045}}+\frac{3\sqrt{2\pi}}{2\lambda^{2.5}}+\frac{7\pi^{4}}{120\lambda^{4}}\right)\;.
\end{equation}
Achieving better accuracy is possible through different algebraic manipulation over the basic structure $\tilde f$, as long as it maintains the correct converges at the edges. To incorporate Padé structure inside the approximation we first manipulate $f_s$ into $\frac{e^{-\lambda}}{\lambda^4}\frac{\frac{9\pi \lambda^{8}}{2}+\frac{49\pi^{8}}{14400}}{1+\lambda^5}$. Then we add powers to complete it into a full Padé structure. By fitting the coefficients we enhance the accuracy as presented in figure~\ref{fig:pressure_less_then_1} by the final expression:
\begin{equation}
\label{eq:mu0decent}
\tilde{f}=\frac{e^{-\lambda}}{\lambda^{4}}\left(\frac{\frac{9\pi \lambda^{8}}{2}+47.95\lambda^{7}+95.3\lambda^{6}+55\lambda^{5}+102.6\lambda^{4}-12\lambda^{3}+37\lambda^{2}+81.1\lambda+\frac{49\pi^{8}}{14400}}{\lambda^{5}-0.33\lambda^{4}+2.18\lambda^{3}-1.597\lambda^{2}+0.522\lambda+1}\right)^{0.5}\;.
\end{equation}

\subsection{Nonzero chemical potential}
\label{sec:nonzero_chemical_potential}

This time we tackle the pressure computation when the chemical potential is not negligible, thus the pressure becomes a function of two variables - temperature and chemical potential. Our strategy is keeping the problem one dimensional by finding a way to approximate $f$ by approximating single variable functions only.

A common way to evaluate $f$, which involves one dimensional expansion only, is Taylor expanding the denominator of eq. \eqref{eq:pressureIntegral} around $\infty$\cite{simplePRDexpansion}:
\begin{equation}
    \frac{1}{e^{x}+1}=\sum_{n=1}^{\infty}{(-1)^{n+1}e^{-nx}}\;.
\end{equation}
By using the modified Bessel function of the second kind integral representation,
\begin{equation}
\label{eq:K2integralrepresentation2}
K_{\alpha}\left(z\right)=\frac{\sqrt{\pi}z^{\alpha}}{2^{\alpha}\Gamma\left(\alpha+\frac{1}{2}\right)}\int_{1}^{\infty}e^{-zt}(t^{2}-1)^{\alpha-\frac{1}{2}}dt\;,
\end{equation}
it is possible to express the pressure as a sum of Bessel functions. This expansion results in fast convergence only for large $1\ll\lambda-\nu=x$. It requires an unbounded number of terms otherwise. To solve that, we replace the common Taylor expansion with an approximation using the presented method. Keeping the approximation analytically integrable, we express the approximation as a sum of exponents only, i.e. $\sum_{n=1}^{N} c_n e^{-a_n x}$.
We choose the range of $0\le x<\infty$, which is sufficient as long as $0 \le m+\mu$. We start by calculating the asymptotic expansions at both range edges:

\begin{equation}
\frac{1}{e^{x}+1} \xrightarrow[x\rightarrow 0]{} \frac12, \quad \frac{1}{e^{x}+1} \xrightarrow[x\rightarrow \infty]{} e^{-x}\;.
\end{equation}

To keep these asymptotic constraints, we take $a_1=c_1=1$, and $1 \leq a_n$. With that expansion the integral takes the form:
\begin{equation}
  f=\sum_{n=1}^{N}\frac{3c_{n}e^{a_{n}\nu}K_{2}\left(-a_{n}\lambda\right)}{a_{n}^{2}\lambda^{2}}\;.
\end{equation}
The coefficients $a_n$, $c_n$, and accuracy for various $N$ values of $\frac{1}{e^{x}+1}$ can be seen in section~\ref{1+x}. With that approach, small number of terms are sufficient for good accuracy across the entire domain. 

\begin{figure}[t!]
 \centering
 \includegraphics[width=0.6\textwidth]{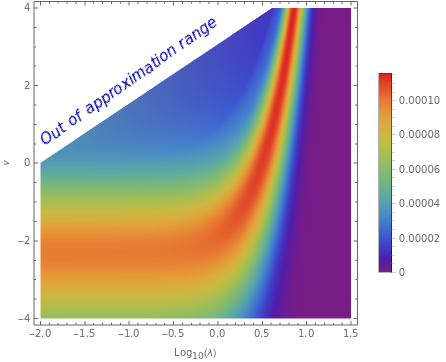}
 \caption{Relative errors for eq. \eqref{eq:general_mu_final} of the Fermi pressure for different temperatures and chemical potentials. The white area is outside of the approximation range of $0\leq m+\mu$.
 } 
\label{fig:IBessels}
\end{figure}

By replacing $\frac{1}{e^{x}+1}$ with a four terms expansion and simplifying , we get the final expression, which is valid for $-m \le \mu$, presented in figure~\ref{fig:IBessels}:
\begin{align}
\label{eq:general_mu_final}
\tilde{f} &= \frac{1}{\lambda^2} \left(3 e^{\nu} K_2(\lambda) - 0.8857 e^{2.0411 \nu} K_2(2.0411 \lambda) \right. \\
 &\quad \left. + 1.0814 e^{2.787 \nu} K_2(2.787 \lambda) - 0.7283 e^{2.92 \nu} K_2(2.92 \lambda)\right)\;.
\end{align}
In case an analytical final form is required, one can substitute $K_2(x)$ by its analytical approximation given in section~\ref{K2}.

\section{Conclusion}
A new procedure for crafting global analytical approximations, accurate over the entire range of the objective function has been demonstrated. They are constructed by first creating asymptotic convergent structures that serve as a scaffold, on top of which additional DOF are blended. These DOF, which usually come in the form of rational functions, control the accuracy of the approximation by numerical minimization. 

First, this procedure has been detailed by directly approximating the pressure of a Fermi gas. This crafted approximation serves as a single expression valid across the entire temperature range, thus making the switch between the known low/high-temperature solutions redundant. Then, we expanded it by considering the chemical potential of a nonzero gas as well. This time, through a slightly different approach, we approximated the integrand of the function. By creating an analytically integrable approximation of the integrand, integration became straightforward.

Finally, this procedure has been utilized to produce global approximations for pivotal functions in physics and cosmology that commonly require evaluation at the transition range between the known asymptotics. We chose the pressure and number density for quantum Bose/Fermi gas and the one-loop correction to the potential in thermal field theory, as we couldn't find a known global approximation for these functions. These functions have been widely discussed in the literature regarding their asymptotic behaviors\cite{simplePRDexpansion} and dedicated programs for their numerical evaluation have been developed\cite{secondPRDexpansions}. Moreover, we apply this procedure to functions for which global approximations were previously discussed, thus improving on previously mentioned results. All plots and approximations presented in this article are available in the following Mathematica notebook\cite{SupplementaryMaterial}.
\appendix

\section{List of Analytical Approximations}
All of the following approximations functions and their residuals are detailed in the supplementary Wolfram Mathematica notebook\cite{SupplementaryMaterial}.

\subsection{Number density of bosons/fermions}

\begin{figure}[t!]

  \centering
  \begin{subfigure}{0.5\textwidth}
    \centering
    \includegraphics[width=\textwidth]{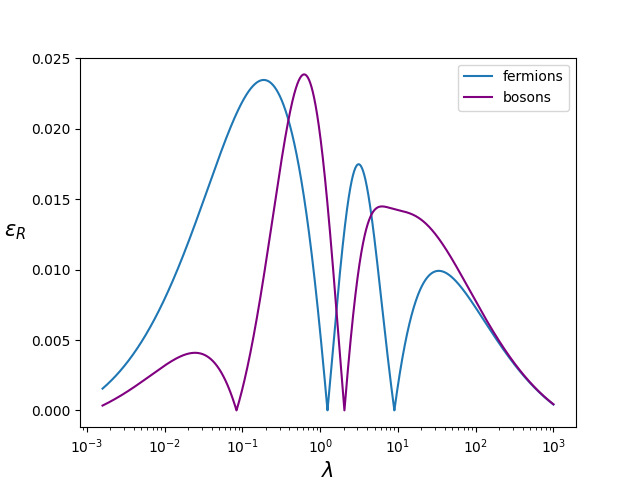}
    \caption{eq.\eqref{eq:app_n1}}.
    \label{fig:n1}
  \end{subfigure}%
  \begin{subfigure}{0.5\textwidth}
    \centering
    \includegraphics[width=\textwidth]{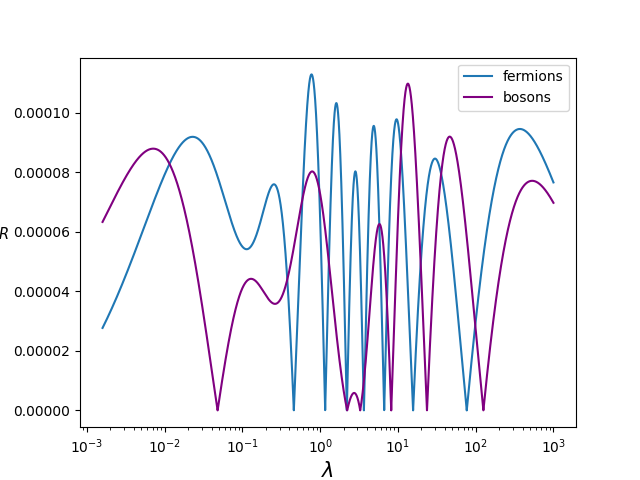}
    \caption{eq.\eqref{eq:app_n1}}.
    \label{fig:n2}
  \end{subfigure}
  \caption{Relative errors, \(\varepsilon_R\), vs. \(\lambda\) for the number density of fermions/bosons.}
  \label{fig:n_density_figs}
\end{figure}

The number density of fermions ($+$) and bosons ($-$), for all energy regimes, from non-relativistic to ultra-relativistic limits, is given by\cite{cosmology}:
\begin{equation}
n=\frac{g}{2\pi^2}T^3I_{\pm}(\lambda)\;,
\end{equation}
where $g$ is the degeneracy factor, $m$ is the mass, $T$ is the temperature, $\lambda=\frac mT$, and
\begin{equation}
I_{\pm}(\lambda)=\int_0^\infty{dx\frac{x^2}{e^{\sqrt{x^2+\lambda^2}}\pm1}}\;.
\end{equation}
We supply two approximations for $I_{\pm}(\lambda)$ over the range $0<\lambda<\infty$, whose accuracy is demonstrated in figure~\ref{fig:n_density_figs}

First Approximation:
\begin{equation}
\label{eq:app_n1}
\tilde I_{\pm}(\lambda)=e^{-\lambda}\left(-\frac{3}{10}\left(\pm1-7\right)+\frac{12}{11}\lambda^{\frac{3}{4}}+\frac{5}{4}\lambda^{\frac{3}{2}}\right)\;.
\end{equation}
Second approximation:
\begin{align}
\label{app_n2}
\tilde{I}_{\pm}(\lambda) &= e^{-\lambda} \left( \sqrt{\frac{\pi}{2}} \lambda^{\frac{3}{2}} + \frac{58}{193} \left(7 \mp 1\right) \right. \\
           &\quad + \frac{1 \pm 1 + 0.155 \lambda^{3} + 1.232 \lambda + \lambda^{0.587 \mp 0.365} \left(\mp 1.024 - 0.135\right) - \lambda^{2.21 \mp 0.313} \left(0.1343 \pm 0.0834\right)}{0.0831 \lambda^{2.464} + 0.81 \lambda^{\mp 0.886 - 0.148} + 1} \Bigg)\;.
\end{align}

\subsection{Pressure and one-loop thermal correction of Bosons/Fermions}
\label{BandFpressureIntegral}
The pressure of Fermi/Bose gas ,$p$\cite{cosmology}, and the one-loop thermal correction $J_{F/B}(\lambda)$\cite{thermalfunctionnumerical}, are both represented by
\begin{equation}
\label{eq:pressureIntegralUnified}
  I_{\pm}=\int_{1}^{\infty}dx\frac{\left(x^{2}-1\right)^{\frac{3}{2}}}{e^{\lambda x}\pm1}\;.
\end{equation}
The pressure of Fermions($+$) and Bosons($-$) gas for all energy regimes, from non-relativistic to ultra-relativistic limits, is given by:
\begin{equation}
   p=\frac{gT^{4}}{6\pi^{2}}\lambda^{4}\times I_{\pm}\;.
\end{equation}
The one-loop thermal correction is given by:
\begin{equation}
J_{F/B}\left(\lambda\right) \equiv  \int_{0}^{\infty} dxx^{2}\ln\left(1 \pm e^{-\sqrt{x^{2}+\lambda^{2}}}\right) = \frac{\lambda^{4}}{3}I_{\pm}\left(\lambda\right)\;,
\end{equation}
where $g$ is the degeneracy factor, $T$ is the temperature, $m$ is the mass and $\lambda=\frac mT$.

We supply an approximation on the range of $0\le \lambda<\infty$, that is more accurate than one percent over the entire range:
\begin{equation}
\label{eq:app_p1}
\tilde{I}_\pm=e^{-\lambda}\left(\frac{3\sqrt{2\pi}}{2\lambda^{2.5}}+\frac{\pi^{4}(15 \mp 1)}{16\cdot15\lambda^{4}}-\frac{1 \pm \frac{1}{5}}{\lambda^{3.6}}+\frac{\frac{22}{5}\pm\frac{3}{20}}{\lambda^{\frac{10}{3}}}\right)\;.
\end{equation}

\begin{figure}[t!]
  \centering
  \includegraphics[width=0.6\textwidth]{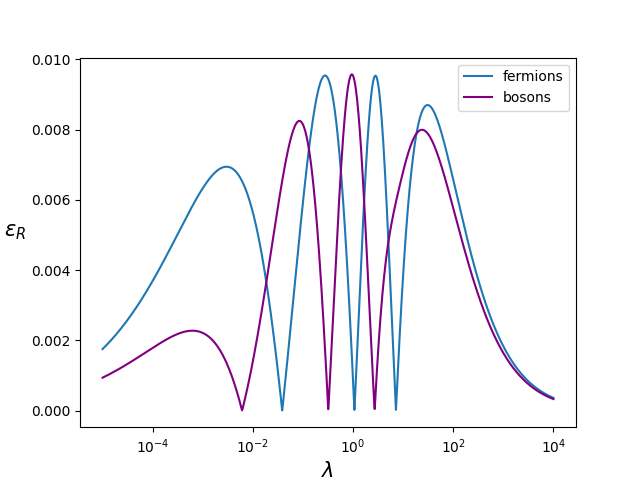}
  \caption{Relative errors, \(\varepsilon_R\), vs. \(\lambda\) for the approximation of Fermi/Bose gas pressure and the one-loop thermal correction at eq. \eqref{eq:app_p1}}.
  \label{fig:JandP1}
\end{figure}

This extends and improves upon the previous results of\cite{ThermalFunctionApproximation}, which produced maximal relative error of around 0.05 for the range of $0 \le \lambda^2 \le 20$.
\subsection{Approximation of \texorpdfstring{$\frac{1}{e^x+1}$}{1/(e\^{}x+1)}}

\label{1+x}
\begin{figure}[t!]
  \centering
   \includegraphics[width=0.6\textwidth]{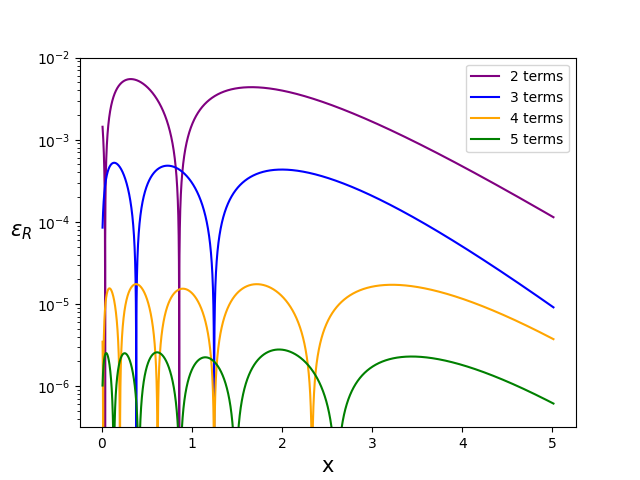}

  \caption{Relative errors, $\varepsilon_R$, as a function of $x$, for eq. \eqref{eq:1+x_app}. 
 } 
 \label{fig:1+x_plots}
\end{figure}

This considered function appears in one form or another in countless integrals. In many cases the following expansion makes it analytically integrable, such as the case of eq. \eqref{eq:pressureIntegral}. We supply four approximations that are presented in figure~\ref{fig:1+x_plots}.
\begin{equation}
\label{eq:1+x_app}
\begin{split}
    & \frac1{e^x+1} \approx -0.502e^{-1.609x}+e^{-x}, \\
    & \frac1{e^x+1} \approx  2.5e^{-2.172x}-3e^{-2.057x}+e^{-x}, \\
    & \frac1{e^x+1} \approx 
 -3.836e^{-2.7392x}+5.1041e^{-2.6425x}-1.76811e^{-2.11001x}+e^{-x}, \\
    & \frac1{e^x+1} \approx 0.60323e^{-3.75592x}-2.30875e^{-3.3734x}+2.2573e^{-3.052x}-1.05178e^{-2.01231x}+e^{-x}\;.
\end{split}
\end{equation}
We note that the value of this function at $x=0$ is $\frac12$. The value of the approximation at $x=0$ is given by $1+\sum_{n=1}^{N} c_n$. By not enforcing this condition we improve the global error on the moderate trade off of nonzero error near $x=0$.

\subsection{Error function}
\begin{figure}[t!]
  \centering
  \includegraphics[width=0.6\textwidth]{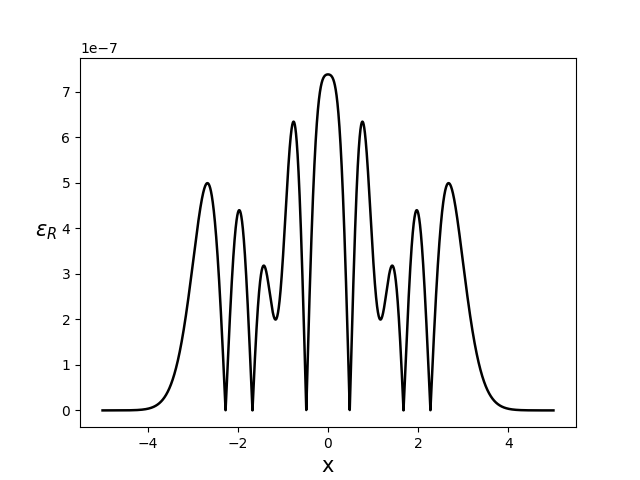}
  \caption{Relative error, \(\varepsilon_R\), vs. \(x\) for the error function approximation of eq. \eqref{eq:erfapp2}.}
  \label{fig:err_app}
\end{figure}

The error function is defined by the integral of a Gaussian:
\begin{equation}
    \mathrm{erf}(x)\equiv\int_{0}^{x}dx e^{-x^{2}}\;.
\end{equation}
We supply an approximation presented in figure~\ref{fig:err_app}, given by:
\begin{equation}
    \label{eq:erfapp2}
    \mathrm{erf} \approx \tanh{\left(1.12838 x \frac{0.007 x^{6} + 0.09126 x^{4} + 0.38336 x^{2} + 1}{0.00162 x^{6} + 0.06485 x^{4} + 0.29228 x^{2} + 1}\right)}\;.
\end{equation}

Another error function approximation that is invertible as well with highest relative error around $2\%$ can be found in\cite{winitzki2003uniform}.

\subsection{The modified Bessel of the second kind \texorpdfstring{$K_2(x)$}{K2(x)}}

\label{K2}
\begin{figure}[t!]
  \centering
  \includegraphics[width=0.6\textwidth]{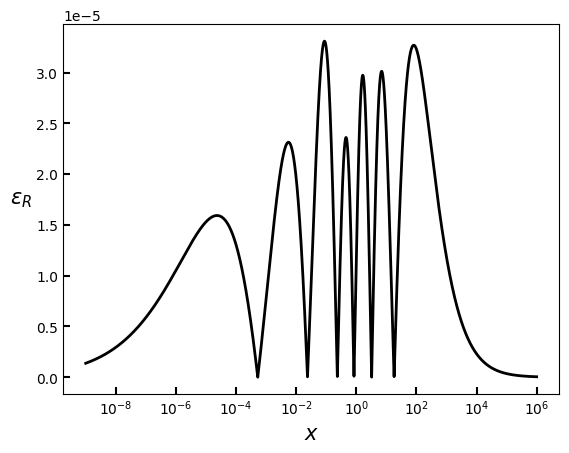}
  \caption{Relative errors, \(\varepsilon_R\), vs. \(x\) for $\tilde{K}_2(x)$ in eq. \eqref{eq:K2model}
 .}
\label{fig:k2}
\end{figure}

The modified Bessel function of the second kind is defined by
\begin{equation}
\label{eq:K2integralrepresentation1}
K_{\nu}\left(z\right)=\frac{\sqrt{\pi}z^{\nu}}{2^{\nu}\Gamma\left(\nu+\frac{1}{2}\right)}\int_{1}^{\infty}e^{-zt}(t^{2}-1)^{\nu-\frac{1}{2}}dt\;.
\end{equation}
We supply an analytical approximation valid for $0\le x<\infty$ for $K_2(x)$ presented in figure~\ref{fig:k2}:
\begin{equation}
\label{eq:K2model}
    \tilde{K}_2=e^{-x}\frac{1+\frac{2.627}{x}+\frac{2}{x^{2}}}{\left(0.033x^{\frac{c_{0}}{5}}-0.102x^{\frac{c_{0}}{4}}+0.113x^{\frac{c_{0}}{3}}+1.162x^{\frac{c_{0}}{2}}+\left(\frac{2x}{\pi}\right)^{c_{0}}+1\right)^{\frac{1}{2c_{0}}}}\;,
\end{equation}
where $c_0=1.984$.

\subsection{PolyLog and Fermi-Dirac integrals}

Among various implications in physics, the PolyLog function $Li_{s}(x)$ relates to the Fermi-Dirac integral by
\begin{equation}
    \frac{1}{\Gamma(s+1)}\int_0^\infty \frac{t^s}{e^{t-x}+1}dt=- Li_{s+1}(- e^x)\;.
\end{equation}
We supply a single approximation that is applicable to important $s$ values, by determining its coefficients: 
\begin{equation}
Li_s(-e^x)\approx-\frac 1{\Gamma(s)} \frac{\Gamma(s)e^{-x}-c_{5}e^{x}+\frac{1}{s}e^{2x}\left(c_{3}+c_{4}x^{2}+c_{6}x^{4}+c_{7}x^{6}+x^{8}\right)^{\frac{s}{8}}+e^{2x}\left(c_{0}+c_{1}x^{2}+c_{2}x^{4}\right)^{c_{13}}}{2\cosh(2x) +\frac{c_{8}+c_{9}x^{2}+c_{10}x^{4}}{c_{11}+c_{12}x^{2}+\frac{x^{4}}{c_{10}}}}\;.
\label{eq:polylog_approximation}
\end{equation}

\begin{table}[]
    \centering
\noindent\makebox[\textwidth][l]{
 \hspace*{-2.2cm} 

 {\small 
 \setlength{\tabcolsep}{3pt} 
\begin{tabular}{llrrrrrrrrrrrrrr}
\hline
 s    & $\| \varepsilon_{R} \|_\infty$   &   $c_{0}$ &   $c_{1}$ &   $c_{2}$ &   $c_{3}$ &   $c_{4}$ &   $c_{5}$ &   $c_{6}$ &   $c_{7}$ &   $c_{8}$ &   $c_{9}$ &   $c_{10}$ &   $c_{11}$ &   $c_{12}$ &   $c_{13}$ \\
\hline
 1/3  & $4.6\cdot10^{-4}$                &    5      &   -1.2    &   0.49    &    663    &   -227    &    3.371  &    139    &   -10.5   &     46    &     236   &       9.7  &     286    &     -10    &   -1.66    \\
 1/2  & $5.9\cdot10^{-4}$                &    7.7    &   -0.67   &   0.057   &    473    &    -23    &    3.376  &     58    &    -7.4   &   -212    &     241   &       9    &     312    &     -11    &   -1.24    \\
 2/3  & $6.5\cdot10^{-4}$                &   11.8    &   -1.1    &   0.09    &    519    &     28    &    3.58   &     57    &    -5.2   &   -547    &     255   &       7.9  &     346    &     -12.7  &   -1.13    \\
 4/3  & $6.9\cdot10^{-5}$                &    1.4732 &    0.125  &   0.00198 &      5.4  &     12.1  &    1.02   &     10.84 &     4.37  &     13    &     135.1 &      11.79 &     404    &     -10.8  &   -1.885   \\
 3/2  & $4.0\cdot10^{-5}$                &    1.52   &    0.0964 &   0.0011  &     15.35 &     28.15 &    1.1931 &     20    &     6.596 &    -94.5  &     131.8 &      11.86 &     443.67 &     -11.46 &   -2.152   \\
 5/3  & $5.8\cdot10^{-5}$                &    3.02   &    0.196  &   0.0093  &     37.8  &     57.3  &    0.9295 &     33.17 &     8.74  &     50.6  &     116   &      12    &     476    &     -11.8  &   -1.283   \\
 2    & $4.7\cdot10^{-5}$                &   24      &   -0.17   &   0.128   &    100.8  &    132    &    0.5685 &     63.5  &    13.14  &    300    &      88.6 &      11.6  &     541    &     -13    &   -0.775   \\
 5/2  & $4.8\cdot10^{-5}$                &   32.2    &   38.7    &  38.8     &      9.75 &     22.3  &    1.206  &     13    &     6.82  &    103.6  &      71.7 &      10    &     651    &     -15.3  &    0.12638 \\
 3    & $2.5\cdot10^{-5}$                &   26.569  &   26.69   &  15.88    &     34.7  &     62.38 &    0.253  &     27.33 &    10.18  &    633.79 &      30.8 &       7.64 &     681.3  &     -17.2  &    0.25076 \\
 7/2  & $3.6\cdot10^{-5}$                &   42.95   &   36.9    &  12.34    &     85.5  &    122.17 &    1.197  &     49.79 &    12.114 &    516.26 &      19.4 &       5.8  &     768.2  &     -20.8  &    0.3759  \\
 4    & $4.6\cdot10^{-5}$                &  103.84   &   92.24   &  19.97    &     26.76 &     17.39 &    7.02   &     13.8  &     3.58  &   -164    &      33   &       5.4  &    1050    &     -27    &    0.50028 \\
 9/2  & $6.4\cdot10^{-5}$                &    9.774  &    2.909  &   0       &    235    &    326    &    0.091  &    126.2  &    15.65  &    715.8  &     -16.4 &       2.6  &     711.8  &     -28    &    1.25    \\
 5    & $2.0\cdot10^{-4}$                &   70.4    &   60.6    &   8.25    &   1855    &    836    &    3      &    179    &    14.3   &    202    &     -19   &       1    &     229    &     -24.6  &    0.7488  \\
 11/2 & $1.4\cdot10^{-4}$                &  149      &   76.3    &   4.63    &   1012    &    887    &  -34.2    &    284    &    28.3   &    342    &     -29.9 &       1.36 &     206    &     -18.8  &    0.8758  \\
 6    & $4.8\cdot10^{-5}$                &  233.98   &  112.86   &   7.98    &     10.3  &     11    &   -1      &      7    &     1.81  &   2004    &     -24   &       2.8  &    1981    &     -30    &    1.00017 \\
 13/2 & $1.2\cdot10^{-4}$                &  280.6    &  122.3    &   5.132   &     22    &     20    &  -82      &      9    &    22.7   &    809    &     -48   &       1.3  &     629    &     -39    &    1.124   \\
\hline
\end{tabular}
}}
    \caption{List of the coefficient values for eq. \eqref{eq:polylog_approximation} corresponding to different $s$ values and their corresponding maximal relative error, denoted as $\|\varepsilon_{R}\|_\infty$ for the PolyLog approximation.}
    \label{tab:polylog_table}
\end{table}
This extends previous work which supplied approximated expressions for $s=\frac12$\cite{nDensity1} for the range of $-2<x<4$ with $\varepsilon_R\approx0.04$ and\cite{nDensity1982} for the range of $-4<x<12$ with $\varepsilon_R\approx0.018$, and the work of\cite{polylogspaper} that approximated various $s$ values in the range $-10<x<10$ with maximal $\varepsilon_R$ varying between $0.01$ and $0.04$.
\subsection{Synchrotron function}
\begin{figure}
    \centering
    \includegraphics[width=0.5\linewidth]{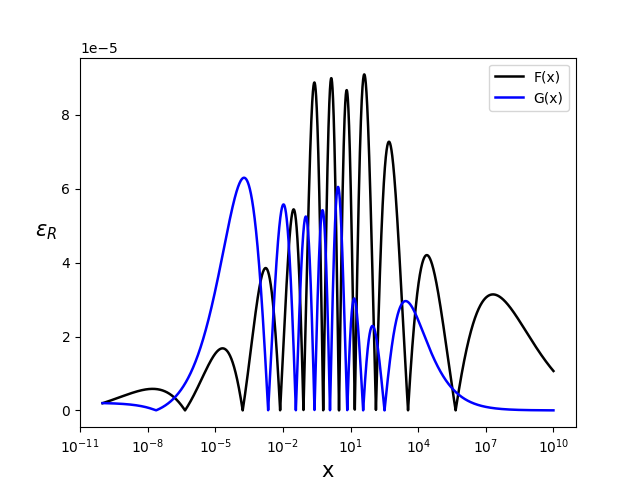}
    \caption{Relative errors, \(\varepsilon_R\), vs. \(x\) for the Synchrotron functions.}
    \label{fig:synchrotronPlot}
\end{figure}

The synchrotron functions, which are used in astrophysics when calculating spectra for different types of synchrotron emission\cite{syncastrobook}, are defined by:
\begin{equation}
    F(x)=x\int_x^\infty K_\frac{5}{3}(t)dt\;,
\end{equation}
\begin{equation}
    G(x)=xK_\frac{2}{3}(x)\;.
\end{equation}
We are approximating them both, on the range of $0\le x<\infty$, as can be seen in figure~\ref{fig:synchrotronPlot}. This extends and improves on the previous results of\cite{synchrotronModels} which produced relative errors of 0.0026 and 0.00035 for $F(x)$ and $G(x)$, respectively.
\begin{equation}
    F(x) \approx e^{-x}\frac{\frac{230}{107}x^{\frac{1}{3}} + x - 0.7078x^{0.701}}{\left(2.835x^{0.61264} - 0.967x^{0.3792} - 0.701x^{0.8239} - 0.61264x^{0.89696} + \left(\frac{2x}{\pi}\right)^{1.1843} + 1\right)^{0.42219}}\;.
\end{equation}
\begin{equation}
    G(x) \approx e^{-x} \frac{\frac{115}{107}x^{\frac{1}{3}} + x}{\left(3.96x^{0.6366} + 4.815x^{1.425} + 1.74x^{2.1711} + \left(\frac{2x}{\pi}\right)^{\frac{54}{19}} + 1\right)^{\frac{19}{108}}}\;.
\end{equation}

\acknowledgments
I would like to deeply thank Anuwedita Singh for her endless conversations and for our engaging discussions. Her support and patience have been invaluable.

I am also profoundly grateful to Yuval Grossman for his guidance. His insights and advice have been crucial in shaping the direction of this work.
\bibliographystyle{JHEP}
\bibliography{biblio.bib}
\end{document}